\documentclass[a4paper,12pt]{article}
\usepackage{mathptmx}
\usepackage[super]{cite}
\usepackage{listings}
\usepackage{geometry}
\usepackage[dvipdfmx]{graphicx,xcolor}
\usepackage{amsmath}
\usepackage{amssymb}
\usepackage{amsthm}
\usepackage{bm}
\usepackage{booktabs}
\usepackage[all]{xy}
\usepackage{multirow}
\usepackage{comment}
\usepackage{threeparttable}
\usepackage{listings}
\usepackage{tabularx}
\usepackage{colortbl}
\usepackage{geometry}
\usepackage[dvipdfmx]{graphicx,xcolor}
\usepackage{amssymb}
\usepackage{mathtools}
\usepackage{ascmac}
\usepackage{fancybox}
\usepackage{latexsym}
\usepackage{mathtools}
\usepackage{amsfonts}
\usepackage{url}
\usepackage{pb-diagram}
\usepackage{color}
\usepackage{autobreak}
\usepackage[normalem]{ulem}
\usepackage{bbm}
\usepackage{setspace}
\usepackage{algorithmic}
\usepackage{algorithm}
\DeclareMathOperator{\Var}{Var}

\DeclareMathOperator{\R}{\mathbb{R}}
\theoremstyle{definition}

\definecolor{mygray}{rgb}{0.9,0.9,0.9}

\title{Asymptotic Properties of Matthews Correlation Coefficient}

\author{Yuki Itaya\thanks{Graduate School of Science and Technology, Keio University, Japan}, \and
        Jun Tamura\thanks{Graduate School of Medicine, Yokohama City University, Japan}, \and
        Kenichi Hayashi\thanks{Department of Mathematics, Keio University, Japan}, \and
        Kouji Yamamoto\thanks{Department of Biostatistics, School of Medicine, Yokohama City University, Japan}}

\begin{document}
\maketitle

\noindent \textbf{Correspondence}\\
Yuki Itaya, Graduate School of Science and Technology, Keio University \\
3-14-1 Hiyoshi, Kohoku, Yokohama, Kanagawa, 223-0061, Japan \\
Email: \url{yuki_0929@keio.jp} \\

\noindent \textbf{Funding information}\\
Funder: Japan Society for the Promotion of Science\\
Grant Number : 21K11790, 23K11013\\


\begin{abstract}
Evaluating classifications is crucial in statistics and machine learning, as it influences decision-making across various fields, such as patient prognosis and therapy in critical conditions. The Matthews correlation coefficient (MCC) is recognized as a performance metric with high reliability, offering a balanced measurement even in the presence of class imbalances. Despite its importance, there remains a notable lack of comprehensive research on the statistical inference of MCC. This deficiency often leads to studies merely validating and comparing MCC point estimates—a practice that, while common, overlooks the statistical significance and reliability of results. Addressing this research gap, our paper introduces and evaluates several methods to construct asymptotic confidence intervals for the single MCC and the differences between MCCs in paired designs. Through simulations across various scenarios, we evaluate the finite-sample behavior of these methods and compare their performances. Furthermore, through real data analysis, we illustrate the potential utility of our findings in comparing binary classifiers, highlighting the possible contributions of our research in this field.
\end{abstract}

\section{Introduction}

Recent advancements in machine learning and big data analytics have significantly increased the importance of classification, not only in the field of medical diagnostics but also across a wide spectrum of other domains. Its use is now crucial  in areas such as cybersecurity threat detection, personalized financial services, and optimizing content recommendations on streaming platforms.
In such diverse contexts, the accurate evaluation of classifier performance is critical. Central to this evaluation is the confusion matrix, a contingency table that evaluates the performance of a classification model. It is a simple yet powerful approach for assessing the effectiveness of a classifier. This matrix compares the actual values with the model's predictions, with each row representing instances in an actual class and each column representing instances in a predicted class (Table \ref{confusion_matrix}). It comprises four elements:
\begin{itemize}
\setlength\itemsep{0.1em}
\item True Positive (TP): Correctly predicted positive observations,
\item False Positive (FP): Incorrectly predicted positive observations,
\item False Negative (FN): Incorrectly predicted negative observations,
\item True Negative (TN): Correctly predicted negative observations.
\end{itemize}
\begin{table}[hbtp]
\begin{center}
\caption{Confusion Matrix.}
\vspace{1mm}
\begin{tabular}{@{}ll|cc@{}}
\toprule
 & & \multicolumn{2}{c}{\textbf{Predicted}} \\
 & & \textbf{Positive} & \textbf{Negative} \\
\midrule
 \multirow{2}{*}{\textbf{Actual:}} & \textbf{Positive} & True Positive (TP) & False Negative (FN) \\
 & \textbf{Negative} & False Positive (FP) & True Negative (TN) \\
\bottomrule
\end{tabular}
\label{confusion_matrix}
\end{center}
\end{table}

Many key performance metrics have been devised based on the confusion matrix. In this paper, we focus on one such performance metric, the Matthews correlation coefficient (MCC), which is defined as
\begin{align}\label{mcc}
\text{MCC} \coloneqq \frac{\text{TP}\times \text{TN}-\text{FP}\times \text{FN}}{\sqrt{(\text{TP}+\text{FP})(\text{TP}+\text{FN})(\text{TN}+\text{FP})(\text{TN} + \text{FN})}}.
\end{align}
MCC is recognized as a robust performance metric, offering a balanced measurement even in the presence of class imbalances. Often referred to as the phi coefficient, MCC ranges between $-1$ and 1, with a value of 1 indicating perfect prediction, 0 indicating random prediction, and $-1$ implying complete disagreement between prediction and actual values. A key characteristic of MCC is that it encompasses all elements of the confusion matrix (TP, FP, FN, and TN). This holistic nature of MCC prevents it from being disproportionately influenced by any single element, thereby rendering it a particularly reliable and informative metric for evaluating the performance of binary classifiers. This is especially crucial in situations where other metrics might present a skewed view, either overly optimistic or pessimistic.\cite{chicco2020advantages, chicco2021matthews, yao2020assessing}

Besides MCC, there are several other performance metrics in practice, the key ones being summarized in Table \ref{various_metrics}. Each metric provides unique insights into classifier performance, and currently, there is no consensus on which is most suitable for specific situations. However, recent studies have highlighted the reliability and informativeness of MCC, implying some kind of superiority to other popular metrics in various contexts.
\begin{center}
\begin{threeparttable}[hbtp]
\caption{Definitions and properties of performance metrics.}
\label{various_metrics}
\begin{tabular}{llp{5cm}p{5cm}}
\toprule
Metric & Definition & Advantages & Disadvantages \\
\midrule \midrule
 Accuracy & $ \frac{\text{TP}+\text{TN}}{\text{TP}+\text{FP}+\text{FN}+\text{TN}} $ & Intuitive and easy to interpret. Clearly represents the proportion of correct predictions. & Not suitable for imbalanced data; it puts more weight on the common classes than on the rare ones \cite{bekkar2013evaluation,gu2009evaluation}.   \\[2mm] \midrule
Balanced accuracy & $\frac{\text{TPR}+\text{TNR}}{2}$\tnote{a} & Effective for imbalanced data. Equally gauges both classes' accuracy. Enables comparison across different datasets\cite{chicco2021matthews, brodersen2010balanced}. &
Can be misleading, as it evaluates a classifier's similarity to random guessing, with a value of 0.5 denoting pure chance \cite{chicco2021matthews}.  \\ \midrule
$\text{F}_1$-score & $2/ \left( \frac{1}{\text{Prec}} + \frac{1}{\text{TPR}} \right)$\tnote{a}  &  Considers the balance between precision and recall. Especially useful when evaluation of TN is relatively unimportant, as it does not account for TN\cite{takahashi2023hypothesis}. & Does not contain information about TN; two distinct classifiers may have the same score. Score is difficult to interpret \cite{chicco2020advantages, yao2020assessing}.  \\ \midrule
MCC & Eq.~\eqref{mcc} & Considers all elements of the confusion matrix, providing an overall quality score. Applicable and reliable even for imbalanced datasets \cite{chicco2020advantages,chicco2021matthews}. & Complicated to comprehend  its asymptotic behavior. Depends on the prevalence\tnote{b}, which hinders making comparisons across different datasets \cite{chicco2021matthews}.\\ \bottomrule
\end{tabular}
\begin{tablenotes}
\footnotesize
\item[a] Here, true positive rate (TPR), true negative rate (TNR), and precision (Prec) are defined as
\[
\text{TPR} \coloneqq \frac{\text{TP}}{\text{TP}+\text{FN}} , \ \ \ \ \text{TNR} \coloneqq \frac{\text{TN}}{\text{TN}+\text{FP}}, \ \ \ \
\text{Prec} \coloneqq \frac{\text{TP}}{\text{TP}+\text{FP}}.
\]
\item[b]  The term ``prevalence" refers to the proportion of truly positive samples within the dataset.\\
\end{tablenotes}
\end{threeparttable}
\end{center}

Chicco et al. conducted a comparative analysis of MCC against other metrics such as accuracy, balanced accuracy, and the $\text{F}_1$-score, revealing several advantages of MCC.\cite{chicco2020advantages,chicco2021matthews} In their studies, they argued that while there are specific scenarios where alternative performance metrics might be preferable, MCC generally emerges as the most informative and reliable due to its comprehensive utilization of all four elements of the confusion matrix. They recommend that practitioners prioritize MCC in most studies, aiming for values close to $+1$. For instance, an MCC of $+0.9$ indicates very high accuracy, balanced accuracy, and $\text{F}_1$-score. In a similar vein, Yao et al. advocated replacing the $\text{F}_1$-score with the unbiased MCC for more accurate performance evaluations.\cite{yao2020assessing} They highlighted the $\text{F}_1$-score's inherent bias, which arises from its ignorance of true negative, and its potential for misleading assessments in cases of imbalanced datasets. The authors also cautioned against the widespread reliance on the $\text{F}_1$-score in software defect detection, noting that its intrinsic biases are frequently ignored, leading to inappropriate applications.

Despite these insights, there remains a notable lack of comprehensive research on the statistical inference of MCC. This gap has led many studies to focus solely on validating and comparing MCC point estimates, a practice that, while common, does not fully address the statistical significance and reliability of the findings.
In light of the need for research dedicated to the statistical inference of MCC, our paper presents and assesses several methods to approximate the asymptotic distribution of MCC, including the differences between MCCs in paired designs. Since MCC is a continuous function of the proportions of TP, FP, FN, and TN relative to the sample size, we apply the Central Limit Theorem---which suggests the convergence of these ratios to their expected values as sample size increases---and the delta method to derive the asymptotic distribution of MCC. Additionally, given that MCC is equivalent to Pearson's correlation coefficient between actual and predicted values \cite{matthews1975comparison}, our research also investigates what insights about MCC can be gleaned  from previous studies on the statistical inference of Pearson's correlation coefficients. While many of these studies typically assumed normality in the underlying distribution\textemdash a factor that might limit direct applications of their theories to MCC\textemdash we have successfully extracted several key ideas and concepts from these previous investigations.

The remainder of this paper is structured as follows: Section 2 discusses two methodologies for constructing asymptotic confidence intervals for a single MCC. Section 3 explores three different approaches for assessing the difference between MCCs in paired designs. Section 4 is dedicated to the validation of these methodologies through simulation studies and an analysis of real-world data. Finally, the paper concludes with a section that summarizes the main findings and discusses potential directions for future research.

\section{Asymptotic distribution of single MCC}

In this section, we introduce two approaches for constructing asymptotic confidence intervals for a single MCC. In the first method, we consider simply applying the delta method to deduce the asymptotic distribution of MCC. For the second, we consider first applying Fisher's $z$ transformation to MCC before following with an application of the delta method. We refer to the first method as the ``Simple method'' and the second as ``Fisher's $z$ method''.
Before delving deeper into these methods, we define some key terms and establish the foundational concepts.

Let $\mathcal{X}$ be an instance space and $\mathcal{Y}=\{0,1\}$ be a label space.
Further, let $\bm{X}$ be a random vector that takes values on $\mathcal{X}$, and $Y$ be a random variable that takes values on $\mathcal{Y}$. Consider a sample $S=\{(\bm{x}_1,y_1),\dots,(\bm{x}_n,y_n)\}$, where each $(\bm{x}_k,y_k)$ is an independent realization of $(\bm{X},Y)$, for $k \in \{1,\dots, n\}$.
Then, for a fixed binary classifier $h: \mathcal{X}\to \mathcal{Y}$, we define $p_{ij}$ and its estimator $\widehat{p}_{ij}$ for $ i,j \in \mathcal{Y}$ as
\begin{align*}
p_{ij} \coloneqq P(h(\bm{X})=i, Y=j), \ \ \ \ \widehat{p}_{ij} \coloneqq \frac{1}{n} \sum_{k=1}^n \mathbbm{1}\{h(\bm{x}_k)=i, y_k=j\}.\end{align*}
Then, we can represent the true positives (TP), false positives (FP), false negatives (FN), and true negatives (TN) as probabilities\footnote{We note that in this context, TP, FP, FN, and TN are conceptualized in probabilistic terms, whereas in Section 1, they are defined based on observations. We assume that such a probabilistic structure underlies the observations in Table 1 of Section 1.}:
\[
\text{TP}=p_{11}, \ \ \text{FP}=p_{10}, \ \ \text{FN}=p_{01}, \ \ \text{TN}=p_{00},
\]
so their estimators can be represented by $\widehat{p}_{ij}$ for $i,j\in \mathcal{Y}$.

For simplicity, we vectorize these quantities by
\[
\bm{p} = (p_{11},p_{10}, p_{01}, p_{00})^{\top} \ \ \ \text{and} \ \ \ \widehat{\bm{p}} = (\widehat{p}_{11},\widehat{p}_{10}, \widehat{p}_{01}, \widehat{p}_{00})^{\top},
\]
where $\top$ denotes transpose.
Here, we define a function $\varphi :(0,1)^4\to [-1,1]$ as
\begin{align}\label{phi}
\varphi(\bm{p})= \frac{p_{11}p_{00}-p_{10}p_{01}}{\sqrt{(p_{11}+p_{10})(p_{11}+p_{01})(p_{00}+p_{10})(p_{00}+p_{01})}},
\end{align}
and then, population and empirical MCCs for the binary classifier $h$ are given by
\[
\text{MCC}[h]= \varphi( \bm{p} )  \ \ \ \text{and} \ \ \ \widehat{\text{MCC}}[h]=\varphi(\widehat{\bm{p}}),
\]
respectively. By the Central Limit Theorem, we have
\begin{align}\label{a1}
\sqrt{n} (\widehat{\bm{p}}-\bm{p}) \rightsquigarrow N(\bm{0},\Sigma(\bm{p})),  \ \ \ \ \text{as} \ n \rightarrow \infty,
\end{align}
where
\begin{align}\label{a4}
\Sigma(\bm{p}) = \begin{pmatrix} p_{11}(1-p_{11}) & -p_{11}p_{10} & -p_{11}p_{01} & -p_{11}p_{00} \\
-p_{10}p_{11} & p_{10}(1-p_{10}) & -p_{10}p_{01} & -p_{10}p_{00} \\
-p_{01}p_{11} & -p_{01}p_{10} & p_{01}(1-p_{01}) & -p_{01}p_{00}\\
-p_{00}p_{11} & -p_{00}p_{10} & -p_{00}p_{01} & p_{00}(1-p_{00})
\end{pmatrix},
\end{align}
because $n\widehat{\bm{p}}$ follows the multinomial distribution with the parameters $n$ and $\bm{p}$.

\subsection{Simple method}
The first approach involves directly applying the delta method to the estimator of MCC in order to obtain its asymptotic distribution.
We call the above the Simple method. 
By the distributional convergence as described in Eq.~\eqref{a1} and the delta method, we have
\[
\sqrt{n}\left( \widehat{\text{MCC}}[h] -\text{MCC}[h] \right) = \sqrt{n} \left( \varphi(\widehat{\bm{p}}) -\varphi(\bm{p}) \right) \rightsquigarrow N(\bm{0}, \nabla \varphi(\bm{p})^{\top} \Sigma(\bm{p}) \nabla \varphi(\bm{p}) ),
\]
where $\nabla$ denotes gradient. Therefore, when the sample size $n$ is large enough, we can approximate the $100(1-\alpha)\%$ confidence interval for $\text{MCC}[h]$ by
\[
\left[ \widehat{\text{MCC}}[h] - z_{\alpha/2} \times \sqrt{\frac{\nabla \varphi(\widehat{\bm{p}})^{\top} \Sigma(\widehat{\bm{p}}) \nabla \varphi(\widehat{\bm{p}})}{n}} ,\ \  \widehat{\text{MCC}}[h] + z_{\alpha/2} \times \sqrt{\frac{\nabla \varphi(\widehat{\bm{p}})^{\top} \Sigma(\widehat{\bm{p}}) \nabla \varphi(\widehat{\bm{p}})}{n}}\right],
\]
where $z_{\alpha/2}$ is the upper $\alpha/2$-quantile of the standard normal distribution.
For reference, the details of $\nabla \varphi (\bm{p})$ is provided in Appendix A.

\subsection{Fisher's $z$ method}
The second approach involves applying Fisher's $z$ transformation to MCC , followed by the application of the delta method.
MCC is equivalent to Pearson's correlation coefficient between actual and predicted values\cite{matthews1975comparison} \hspace{-1mm}; that is,
\[
\text{MCC}[h]= \frac{\text{Cov}(h(\bm{X}),Y)}{\sqrt{\Var(h(\bm{X}))\Var(Y)}}.
\]
Thus, by drawing upon the asymptotic theory related to Pearson's correlation coefficient, we analogously apply Fisher's $z$ transformation to MCC, expecting a more rapid convergence towards a normal distribution. Fisher's $z$ transformation utilizes the inverse hyperbolic tangent (artanh) to transform the correlation coefficient, expanding its range from $(-1, 1)$ to $(-\infty, \infty)$. This adjustment mitigates the skewness of the sample correlation coefficient's distribution, rendering it more close to a normal distribution.

The sampling distributions for MCC are often highly skewed, requiring both large sample sizes and middle-sized correlations for the Simple method to be accurate, that is, to provide adequate coverage in a non-lopsided fashion. On the other hand, Fisher's $z$ transformation for Pearson's correlation coefficient is known to perform very well \cite{lee1971some}. The accuracy of these procedures originates largely from respecting the asymmetric feature  of the sampling distributions \cite{zou2007toward}. Given the context described above, in this subsection, we derive the asymptotic confidence interval for a single MCC by applying Fisher's $z$ transformation.

Define the function $f:(-1,1)\to \R$ as
\[
f(x)\coloneqq \text{artanh}(x) = \frac{1}{2}\log\left( \frac{1+x}{1-x} \right).
\]
Let $r_n$ represents the sample correlation coefficient calculated from iid sample $(X_i,Y_i)_{i=1}^n$. It is well known that if $(X_i,Y_i)$ are normally distributed, the asymptotic variance of $f(r_n)$ is given by $1/(n-3)$ \cite{olkin1976asymptotic}. However, in the context of MCC, this assumption does not hold. Therefore, to derive the asymptotic variance of MCC, we employ the delta method. By the distributional convergence in Eq.~\eqref{a1} and the delta method, we have
\[
\sqrt{n} (f(\widehat{\text{MCC}}[h])-f(\text{MCC}[h])) = \sqrt{n} (f\circ \varphi(\widehat{\bm{p}})-f\circ \varphi(\bm{p})) \rightsquigarrow N \left(\bm{0}, \{\nabla f\circ \varphi (\bm{p})\}^{\top} \Sigma(\bm{p}) \{ \nabla f\circ \varphi (\bm{p})\} \right).
\]
Therefore, the lower bound $L$ and the upper bound $U$ of the asymptotic $100(1-\alpha)\%$ confidence interval for $f(\text{MCC}[h])$ are given by
\[
(L,U) = f(\widehat{\text{MCC}}[h]) \mp z_{\alpha/2} \times \sqrt{\frac{\{\nabla f\circ \varphi(\widehat{\bm{p}})\}^{\top} \Sigma(\widehat{\bm{p}}) \{\nabla f\circ \varphi(\widehat{\bm{p}})\}}{n}}.
\]
By the strict monotonicity of the function $f$, the asymptotic $100(1-\alpha)\%$ confidence interval for $\text{MCC}[h]$ is given by
\[
\left[ f^{-1}(L) ,  f^{-1}(U)\right].
\]

\section{Asymptotic distribution of MCC difference in paired design}
In practical scenarios involving the evaluation of multiple classifiers, it is often crucial to determine whether a statistically significant difference exists between two MCCs. When two MCCs are independent, assessing the statistical significance of their difference becomes a more straightforward process. That is, we are able to derive the asymptotic distribution for the difference between two MCCs by applying the approach for the single case to each MCC. However, when two MCCs are not independent, deriving the asymptotic distribution of their difference is not as straightforward. In practical applications, there are many situations where the two MCCs are not independent, especially in the form of  comparisons in paired design where each subject is assessed by two different classifiers.

The paired design approach effectively mitigates variability arising from individual differences, making it invaluable across various research fields. For example, consider a scenario where two screening tests are conducted on the same group of subjects. By analyzing MCC differences, researchers can pinpoint the more consistent and effective screening test---a vital process in fields like medicine, where accurate diagnosis can significantly impact patient outcomes. Considering the significant implication of  comparing two MCCs in paired designs, this section is dedicated to comparing two correlated MCCs derived from different binary classifiers applied to the same dataset.

To further elaborate, it is essential to distinguish between single MCC analysis and the comparison of MCCs in paired designs. In the preceding section, we obtained the confidence interval for a single MCC by using Fisher's $z$ transformation. However, this approach is not directly applicable for $\text{MCC}[h_1]-\text{MCC}[h_2]$, where $h_1$ and $h_2$ are two distinct classifiers. This is because the confidence limits for $f(\text{MCC}[h_1])-f(\text{MCC}[h_2])$ cannot be back-transformed to obtain the interval for $\text{MCC}[h_1]-\text{MCC}[h_2]$ \cite{meng1992comparing,olkin1995correlations}. In response to this challenge, alongside the Simple method, we explore two additional approaches: one proposed by Zou \cite{zou2007toward}, and another which applies a transformation---a slightly modified version of Fisher's $z$ transformation---to the difference between MCCs. We refer to these methods as ``Zou's method" and the ``Modified Transformation (MT) method", respectively.

For fixed binary classifiers $h_1,h_2: \mathcal{X}\to \mathcal{Y}$, we define
\begin{align}\label{a2}
p_{ijk} \coloneqq P(h_1(\bm{X})=i, h_2(\bm{X})=j, Y=k) \ \ \ \ \text{for} \ \ i,j,k \in \mathcal{Y}
\end{align}
Then, using $p_{ijk}$'s in Eq.~\eqref{a2} , we can represent TP, FP, FN, and TN for classifier $h_1$ as
\begin{align*}
 \text{TP}[h_1] =P(h_1(\bm{X})=1, Y=1) =  p_{111}+p_{101},  \ \ \ \ \ & \text{FP}[h_1]  = P(h_1(\bm{X})=1, Y=0) = p_{110}+p_{100}, \\
\text{FN}[h_1] = P(h_1(\bm{X})=0, Y=1) = p_{011}+p_{001},  \ \ \ \ \ & \text{TN}[h_1]  = P(h_1(\bm{X})=0, Y=0) = p_{010}+p_{000},
\end{align*}
and TP, FP, FN, and TN for classifier $h_2$ as
\begin{align*}
\text{TP}[h_2] =P(h_2(\bm{X})=1, Y=1) =  p_{111}+p_{011},  \ \ \ \ \ &  \text{FP}[h_2]  = P(h_2(\bm{X})=1, Y=0) = p_{110}+p_{010}, \\
\text{FN}[h_2] = P(h_2(\bm{X})=0, Y=1) = p_{101}+p_{001},  \ \ \ \ \ & \text{TN}[h_2]  = P(h_2(\bm{X})=0, Y=0) = p_{100}+p_{000}.
\end{align*}
Thus, using the function $\varphi$ defined in Eq.~\eqref{phi}, MCC for each classifier is given by
\[
\text{MCC}[h_l]=\varphi(\text{TP}[h_l],\text{FP}[h_l],\text{FN}[h_l],\text{TN}[h_l]) \ \ \ \ \text{for} \ \  l=1,2.
\]
Let the estimator of $p_{ijk}$, denoted by $\widehat{p}_{ijk}$, be the empirical probability, just as in the single case. By plugging in these estimators, we can calculate the estimators of TP, FP, FN, and TN for each classifier, denoted with a hat. Subsequently, the estimator of MCC for each classifier is expressed as
\[
\widehat{\text{MCC}}[h_l] = \varphi(\widehat{\text{TP}}[h_l], \widehat{\text{FP}}[h_l], \widehat{\text{FN}}[h_l], \widehat{\text{TN}}[h_l])\ \ \ \ \text{for} \ \  l=1,2.
\]
For simplicity, let $\bm{p}$ and $\widehat{\bm{p}}$ be the eight-dimensional column vectors whose components are $p_{ijk}$ and $\widehat{p}_{ijk}$, respectively, for all possible combinations of $i,j,k \in \mathcal{Y}$.
Then, in the same way as in the single case, by the Central Limit Theorem, we have
\begin{align}\label{a3}
\sqrt{n} (\widehat{\bm{p}}-\bm{p}) \rightsquigarrow N(0,\Sigma(\bm{p})),
\end{align}
where $n$ indicates the sample size, and the asymptotic variance $\Sigma(\bm{p})$ is obtained in the same way as in the single case considering that $n\widehat{\bm{p}}$ follows the eight-dimensional multinomial distribution.

\subsection{Simple method for paired design}
This approach is essentially the same as the Simple method for the single case.
Define the function $\psi : (0,1)^8 \to [-2,2]$ as

\begin{align}
\begin{split}
\psi(\bm{p}) &\coloneqq \varphi(p_{111}+p_{101},p_{110}+p_{100},p_{011}+p_{001},p_{010}+p_{000}) \\ &\hspace{5mm} -\varphi(p_{111}+p_{011},p_{110}+p_{010},p_{101}+p_{001},p_{100}+p_{000}) \end{split} \label{psi}\\
&= \text{MCC}[h_1]-\text{MCC}[h_2] \notag
\end{align}
and then,
\[
\psi (\widehat{\bm{p}}) = \widehat{\text{MCC}}[h_1] - \widehat{\text{MCC}}[h_2].
\]
Therefore, by the distributional convergence \eqref{a3} and the delta method, we have
\[
\sqrt{n} (\psi(\widehat{\bm{p}})-\psi(\bm{p})) \rightsquigarrow N(0, \nabla \psi(\bm{p})^{\top} \Sigma(\bm{p}) \nabla \psi(\bm{p})  ).
\]
This allows us to derive the asymptotic confidence interval for $\psi(\bm{p})=\text{MCC}[h_1]-\text{MCC}[h_2]$.
For reference, the result of the differentiation of $\psi(\bm{p})$ is provided in Appendix A.

\subsection{Zou's method}
As mentioned earlier, an alternative to Fisher's $z$ method needs to be considered because directly applying Fisher's $z$ transformation  to each individual MCC does not yield the asymptotic confidence interval for the difference between MCCs. Zou (2007) introduced a method that accounts for the asymmetry and skewness in the sampling distribution of the correlation coefficient\cite{chicco2021matthews}. In his research, however, Zou assumed normality of the underlying distribution and used an approximation formula that was based on the findings of Olkin and Shiotani (1976) to calculate the correlation between two sample correlation coefficients\cite{olkin1976asymptotic}. Consequently, since this approximation is not directly applicable to our case, we utilize the delta method to approximate the correlation coefficient between two MCCs.

In Zou's method, we first calculate the asymptotic $100(1-\alpha)\%$ confidence interval for each MCC by Fisher's $z$ method. For $l=1,2$, let $\ell_l$ and $u_l$ represent the lower and upper bounds of these intervals for $\text{MCC}[h_l]$, respectively.
Using these $\ell_l$ and $u_l$ for $l=1,2$ and the notation $\widehat{r}_1=\widehat{\text{MCC}}[h_1]$ and $\widehat{r}_2=\widehat{\text{MCC}}[h_2]$ for simplicity, the asymptotic $100(1-\alpha)\%$ confidence lower limit $L$ and upper limit $U$ are given by
\begin{align*}
\begin{cases}
L = \widehat{r}_1-\widehat{r}_2  - \sqrt{(\widehat{r}_1-\ell_1)^2+(u_2-\widehat{r}_2)^2 -2 \widehat{\text{Corr}}(\widehat{r}_1,\widehat{r}_2)(\widehat{r}_1-\ell_1)(u_2-\widehat{r}_2)} \\
U =  \widehat{r}_1-\widehat{r}_2  + \sqrt{(u_1-\widehat{r}_1)^2+(\widehat{r}_2-\ell_2)^2 -2 \widehat{\text{Corr}}(\widehat{r}_1,\widehat{r}_2)(u_1-\widehat{r}_1)(\widehat{r}_2-\ell_2)} ,
\end{cases}
\end{align*}
where $\widehat{\text{Corr}}(\widehat{r}_1, \widehat{r}_2)$ is the estimated correlation between the two MCCs, calculated as
\begin{align*}
\widehat{\text{Corr}}(\widehat{r}_1, \widehat{r}_2)
= \widehat{\text{Cov}}(\widehat{r}_1,\widehat{r}_2)/ \sqrt{\widehat{\Var}(\widehat{r}_1)\widehat{\Var}(\widehat{r}_2)}.
\end{align*}
To estimate the variance matrix of $\left( \widehat{r}_1,\widehat{r}_2 \right)^{\top}$, define the function $\tilde{\psi} : (0,1)^8 \to [-1,1]^2 $ as
\[
\tilde{\psi} (\bm{p}) \coloneqq \begin{pmatrix} \varphi(p_{111}+p_{101},p_{110}+p_{100},p_{011}+p_{001},p_{010}+p_{000}) \\\varphi(p_{111}+p_{011},p_{110}+p_{010},p_{101}+p_{001},p_{100}+p_{000}) \end{pmatrix},
\]
and then, by the distributional convergence \eqref{a3} and the delta method,
\[
\sqrt{n} \left(\begin{pmatrix} \widehat{r}_1 \\ \widehat{r}_2 \end{pmatrix} - \begin{pmatrix}\text{MCC}[h_1] \\ \text{MCC}[h_2] \end{pmatrix}\right) = \sqrt{n} \left( \tilde{\psi}(\widehat{\bm{p}})- \tilde{\psi}(\bm{p}) \right) \rightsquigarrow N\left( \bm{0}, \nabla \tilde{\psi}(\bm{p})^{\top} \Sigma(\bm{p}) \nabla \tilde{\psi}(\bm{p}) \right).
\]
By plugging in the estimators of $\bm{p}$ to the limiting variance above, we can estimate the variance matrix of $\left( \widehat{r}_1,\widehat{r}_2 \right)^{\top}$.

\subsection{Modified Transformation (MT) method }
In the third approach, we consider applying Fisher's $z$ transformation to half the difference between MCCs. This approach enables us to account for the skewness observed in the distribution of the difference between MCCs. Define the function $g:(-2,2)\to \R$ as
\[
g(x)=\frac{1}{2} \log \left( \frac{2+x}{2-x} \right)
\]
and then, by the distributional convergence \eqref{a3} and the delta method,
\begin{align*}
\sqrt{n} \left( g(\widehat{\text{MCC}}[h_1] - \widehat{\text{MCC}}[h_2]) -g(\text{MCC}[h_1]-\text{MCC}[h_2]) \right)
&= \sqrt{n} \left( g\circ \psi(\widehat{\bm{p}}) - g\circ \psi(\bm{p})  \right) \\
&\rightsquigarrow N\left(\bm{0}, \{\nabla g\circ \psi (\bm{p})\}^{\top} \Sigma(\bm{p}) \{ \nabla g\circ \psi (\bm{p})\}\right).
\end{align*}
Therefore, the lower bound $L$ and the upper bound $U$ of the asymptotic $100(1-\alpha)\%$ confidence interval for $g(\text{MCC}[h_1]-\text{MCC}[h_2])$ are given by
\[
(L,U) = g(\widehat{\text{MCC}}[h_1] - \widehat{\text{MCC}}[h_2]) \mp z_{\alpha/2} \times \sqrt{\frac{\{\nabla g\circ \psi (\widehat{\bm{p}})\}^{\top} \Sigma(\widehat{\bm{p}}) \{\nabla g\circ \psi (\widehat{\bm{p}})\}}{n}}.
\]
By the strict monotonicity of the function $g$, the asymptotic $100(1-\alpha)\%$ confidence interval for $\text{MCC}[h_1]-\text{MCC}[h_2]$ is given by
\[
\left[ g^{-1}(L), g^{-1}(U) \right].
\]

\section{Numerical studies}
In this section, through simulations across various scenarios, we evaluate the finite-sample behavior of the methods introduced in this paper, and also compare the performances of the methods. Furthermore, through real data analysis, we validate the practicality of our findings. In order to help to reproduce the results and calculate confidence intervals of MCC, we provide the  R source codes freely at \url{https://github.com/yukiitaya/MCC}.

In our simulation study, we construct 95\% confidence intervals $m=1,000,000$ times to estimate the coverage probability. It is important to note that when the sample size is small, instances occur where we cannot define $\widehat{\text{MCC}}$ as $\widehat{\text{TP}}+\widehat{\text{FN}}=0$. Moreover, there are instances where we cannot apply Fisher's $z$ transformation to $\widehat{\text{MCC}}$ as it equals 1. Under such circumstances, deriving the confidence interval for MCC is not possible. Hence, throughout the $m$ trials, these particular cases are omitted in estimating the coverage probability of the confidence interval. 

\subsection{Simulation study for single MCC }
We explore various scenarios by changing the values of $P(Y=1)$ (0.1 for imbalanced situations, 0.5 for balanced situations) and the true MCC (0.4, 0.6, or 0.8), with the condition $\text{TP}/\text{FN} = \text{TN}/\text{FP}$. The specific values of TP, FP, FN, and TN for each case are provided in Appendix B. The simulation results are shown in Table \ref{single}. We also report the coverage probabilities, assuming that the asymptotic variances of the Fisher-transformed MCCs are given by $1/(n-3)$. If there are cases where calculating the confidence interval for the MCC is infeasible, the number of such cases is also displayed next to each coverage probability estimate.

Table \ref{single} demonstrates that Fisher's $z$ method outperforms the Simple method in deriving the asymptotic confidence intervals for a single MCC. It is also evident that approximating the asymptotic variances of the Fisher-transformed MCCs by $1/(n-3)$ is not appropriate.
Additionally, it is observed that in balanced situations, MCC estimates are more accurately approximated by a normal distribution compared to those in imbalanced situations. Moreover, the performance of the approximation tends to decrease as the MCC value moves away from zero.

\begin{table}[hbtp]
\centering
\setlength{\tabcolsep}{10pt}
\caption{Estimated coverage probability for single MCC.}
\begin{tabular}{llllll}
\toprule
 $P(Y=1)$ & MCC   & $n$ & Simple (NA) & Fisher's $z$ (NA) & var $\approx \frac{1}{n-3} $ (NA) \\
\midrule
 0.1 & 0.4 & 50 & .9148 (5170)  & \textbf{.9366} (5180) &.8816 (5180) \\
 \rowcolor{mygray!70}   &  & 100 & .9317 (33) & \textbf{.9415} (33) &.8858 (33) \\
 &  & 500 & .9467 & \textbf{.9483} & .8864 \\
 \rowcolor{mygray!70}   &  & 1000 & .9485 & \textbf{.9494} & .8868 \\
 &  & 5000 & \textbf{.9500} & .9502 & .8872 \\
 \rowcolor{mygray!70}   &  & 10000 & .9499 & \textbf{.9500} & .8868 \\
\hline
 & 0.6 & 50 & .9130 (5209)  & \textbf{.9538} (8194) & .7668 (8194) \\
 \rowcolor{mygray!70}   &  & 100 & .9311 (30) & \textbf{.9513} (37) & .7741 (37) \\
 &  & 500 & .9466 & \textbf{.9499} & .7785  \\
 \rowcolor{mygray!70}   &  & 1000 & .9482 & \textbf{.9499} & .7789 \\
 &  & 5000 & .9497  & \textbf{.9501} & .7790 \\
 \rowcolor{mygray!70}   &  & 10000 & .9499 & \textbf{.9500} & .7789 \\
\hline
 & 0.8 & 50 & .8581 (5258)  & \textbf{.9406} (109812) & .6231 (109812)  \\
 \rowcolor{mygray!70}   &  & 100 & .9061 (28) & \textbf{.9604} (9496) & .5752 (9496)  \\
  &  & 500 & .9414  & \textbf{.9513} & .5712 \\
 \rowcolor{mygray!70}   &  & 1000 & .9459 & \textbf{.9509} & .5773  \\
 &  & 5000 & .9493  & \textbf{.9502} & .5779  \\
 \rowcolor{mygray!70}   &  & 10000 & .9493 & \textbf{.9498} & .5784 \\
\midrule

 0.5 & 0.4 & 50 & .9351  & \textbf{.9520} & .9301  \\
 \rowcolor{mygray!70}   &  & 100 & .9428 & \textbf{.9518} & .9238  \\
 &  & 500 & \textbf{.9487}  & .9514 & .9269  \\
 \rowcolor{mygray!70}  &  & 1000 & \textbf{.9499} & .9508 & .9276 \\
 &  & 5000 & \textbf{.9501}  & .9503 & .9273  \\
 \rowcolor{mygray!70}  &  & 10000 & \textbf{.9497} & .9495 & .9272  \\
\hline
 & 0.6 & 50 & .9276  & \textbf{.9504} (13)  & .8677 (13)  \\
 \rowcolor{mygray!70}  &  & 100 & .9353 & \textbf{.9473} & .8792 \\
 &  & 500 & .9480  & \textbf{.9498} & .8826 \\
 \rowcolor{mygray!70}   &  & 1000 & .9480 & \textbf{.9488} & .8841 \\
 &  & 5000 & \textbf{.9502}  & .9504 & .8836 \\
 \rowcolor{mygray!70}   &  & 10000 & .9497 & \textbf{.9500} & .8829  \\
\hline
 & 0.8 & 50 & .8845 (784)  & \textbf{.9718} (4870) & .7972 (4870)  \\
\rowcolor{mygray!70}  &  & 100 & .9304 (3) & \textbf{.9553} (29) & .7796 (29)  \\
&  & 500 & .9423  & \textbf{.9539} & .7638  \\
\rowcolor{mygray!70}  &  & 1000 & .9508 & \textbf{.9496} &  .7548 \\
&  & 5000 & .9507  & \textbf{.9498} & .7609 \\
\rowcolor{mygray!70} &  & 10000 & \textbf{.9503} & .9491 & .7577  \\
\bottomrule
\end{tabular}\\
Note: Values closest to 0.95 are displayed in bold.
\label{single}
\end{table}

\subsection{Simulation study for MCC difference in paired design }
We explore various scenarios by changing the values of $P(Y=1)$ (0.1 for imbalanced situation, 0.5 for balanced situation) and the true MCCs (0.4, 0.6, or 0.8)\footnote{To uniquely identify $p_{ijk}$ for $i,j,k \in \mathcal{Y}$ defined in Eq.~\eqref{a2}, we additionally set four constant conditions: $p_{001}=0.001$, $p_{110}=0.01$, $\text{TP}_1/\text{FN}_1 = \text{TN}_1/\text{FP}_1$, $\text{TP}_2/\text{FN}_2= \text{TN}_2/\text{FP}_2$.}.
The specific values of $\text{TP}_l$, $\text{TN}_l$, $\text{FP}_l$, and $\text{FN}_l$ for $l=1,2$ corresponding to each scenario are provided in Appendix B. The simulation results are shown in Tables \ref{paired1} and \ref{paired2}. If there are cases where calculating the confidence interval for the MCC difference is infeasible, the number of such cases is also displayed next to each coverage probability estimate.

Tables \ref{paired1} and \ref{paired2} suggest that both Zou's method and the MT method outperform the Simple method. However, determining the superiority between Zou's method and the MT method is not straightforward. While the MT method consistently approaches the nominal 95\% level from below, Zou's method tends to yield conservative results, as the sample size is small and MCCs are high. As with single cases, it can be observed that in balanced situations, the coverage probability approaches the nominal value more rapidly compared to imbalanced situations.
\begin{table}[hbtp]
\centering
\caption{Estimated coverage probability for MCC difference in paired design (imbalanced).}
\setlength{\tabcolsep}{10pt}
\begin{tabular}{lllllll}
\toprule
$P(Y=1)$ & $\text{MCC}_1$ & $\text{MCC}_2$ & $n$ & Simple (NA) & Zou's (NA) & MT (NA)  \\
\midrule
0.1 & 0.4 & 0.4 & 50 & .9040 (5201)  & \textbf{.9181} (5947) & .9117 (5201)  \\
\rowcolor{mygray!70}  &  & & 100 & .9276 (37) & .9305 (38) & \textbf{.9306} (37)  \\
&  & & 500 & .9459 & .9463 & \textbf{.9466}  \\
 \rowcolor{mygray!70}  &  & & 1000 & .9483 & .9485 & \textbf{.9486}  \\
 &  & & 5000 & \textbf{.9502} & \textbf{.9502} & .9503 \\
 \rowcolor{mygray!70}  &  & & 10000 & .9499 & .9499 & \textbf{.9500}  \\
\hline
 &  & 0.6 & 50 & .9092 (5209)  & \textbf{.9297} (7470) & .9156 (5209)  \\
\rowcolor{mygray!70}  &  & & 100 & .9304 (26) & \textbf{.9372} (34) & .9332 (26)  \\
&  & & 500 & .9467 & \textbf{.9482} & .9473  \\
 \rowcolor{mygray!70}  &  & & 1000 & .9485 & \textbf{.9495} & .9488  \\
 &  & & 5000 & .9497 & \textbf{.9501} & .9498  \\
 \rowcolor{mygray!70}  &  & & 10000 & .9498 & \textbf{.9500} & .9498  \\
\hline
 &  & 0.8 & 50 & .9028 (5215)  & \textbf{.9110} (85631) & .9063 (5215)  \\
\rowcolor{mygray!70}  &  & & 100 & .9295 (18) & \textbf{.9427} (9536) & .9318 (18)  \\
&  & & 500 & .9465 & \textbf{.9512} & .9469  \\
 \rowcolor{mygray!70}  &  & & 1000 & .9482 & \textbf{.9512} & .9485  \\
 &  & & 5000 & \textbf{.9499} & .9511 & \textbf{.9499}  \\
 \rowcolor{mygray!70}  &  & & 10000 & \textbf{.9496} & \textbf{.9504} & \textbf{.9496}  \\
 \hline
 & 0.6 & 0.6 & 50 & .9234 (5202)  & .9731 (16214) & \textbf{.9282} (5202)  \\
\rowcolor{mygray!70}  &  & & 100 & .9385 (22) & \textbf{.9571} (128) & .9413 (22)  \\
&  & & 500 & .9474 & \textbf{.9498} & .9479  \\
 \rowcolor{mygray!70}  &  & & 1000 & .9488 & \textbf{.9500} & .9490  \\
 &  & & 5000 & .9497 & \textbf{.9499} & .9497  \\
 \rowcolor{mygray!70}  &  & & 10000 & \textbf{.9500} & .9501 & \textbf{.9500}  \\
  \hline
 & & 0.8 & 50 & .9186 (5251)  & \textbf{.9614} (89304) & .9226 (5251)  \\
\rowcolor{mygray!70}  &  & & 100 & .9368 (29) & .9667 (9508) & \textbf{.9388} (29)  \\
&  & & 500 & .9476 & .9555 & \textbf{.9480}  \\
 \rowcolor{mygray!70}  &  & & 1000 & .9487 & .9537 & \textbf{.9489}  \\
 &  & & 5000 & \textbf{.9497} & .9514 & \textbf{.9497}  \\
 \rowcolor{mygray!70}  &  & & 10000 & .9498 & .9510 & \textbf{.9499}  \\
\hline
 & 0.8 & 0.8 & 50 & .9032 (5318)  & \textbf{.9422} (150903) & .9052 (5318)  \\
\rowcolor{mygray!70}  &  & & 100 & .9333 (30) & .9891 (18739) & \textbf{.9339} (30)  \\
&  & & 500 & .9476 & .9587 & \textbf{.9479}  \\
 \rowcolor{mygray!70}  &  & & 1000 & .9489 & .9543 & \textbf{.9491}  \\
 &  & & 5000 & \textbf{.9500} & .9511 & \textbf{.9500}  \\
 \rowcolor{mygray!70}  &  & & 10000 & \textbf{.9499} & .9504 & \textbf{.9499}  \\
\bottomrule
\end{tabular}\\
Note: Values closest to 0.95 are displayed in bold.
\label{paired1}
\end{table}

\begin{table}[hbtp]
\centering
\caption{Estimated coverage probability for MCC difference in paired design (balanced).}
\setlength{\tabcolsep}{10pt}
\begin{tabular}{lllllll}
\toprule
$P(Y=1)$ & $\text{MCC}_1$ & $\text{MCC}_2$ & $n$ & Simple (NA) & Zou's (NA) & MT (NA)  \\
\midrule
0.5 & 0.4 & 0.4 & 50 & .9374  & .9442 & \textbf{.9444}  \\
\rowcolor{mygray!70}  &  & & 100 & .9442 & .9473 & \textbf{.9476}  \\
&  & & 500 & .9487 & .9493 & \textbf{.9494}  \\
 \rowcolor{mygray!70}  &  & & 1000 & \textbf{.9499} & .9502 & .9503  \\
 &  & & 5000 & .9495 & \textbf{.9496} & \textbf{.9496}  \\
 \rowcolor{mygray!70}  &  & & 10000 & .9497 & \textbf{.9498} & \textbf{.9498}  \\
\hline
 &  & 0.6 & 50 & .9378  & \textbf{.9450} (8) & .9435  \\
\rowcolor{mygray!70}  &  & & 100 & .9443 & .9471 & \textbf{.9472}  \\
&  & & 500 & .9489 & .9491 & \textbf{.9495}  \\
 \rowcolor{mygray!70}  &  & & 1000 & .9498 & .9496 & \textbf{.9500}  \\
 &  & & 5000 & \textbf{.9498} & .9497 & \textbf{.9498}  \\
 \rowcolor{mygray!70}  &  & & 10000 & \textbf{.9498} & .9497 & \textbf{.9498}  \\
\hline
 &  & 0.8 & 50 & .9377  & \textbf{.9527} (4424) & .9420  \\
\rowcolor{mygray!70}  &  & & 100 & .9441 & \textbf{.9492} (18) & .9463  \\
&  & & 500 & .9490 & .9493 & \textbf{.9494}  \\
 \rowcolor{mygray!70}  &  & & 1000 & .9496 & .9495 & \textbf{.9498}  \\
 &  & & 5000 & .9499 & .9497 & \textbf{.9500}  \\
 \rowcolor{mygray!70}  &  & & 10000 & \textbf{.9498} & .9496 & \textbf{.9498}  \\
 \hline
 & 0.6 & 0.6 & 50 & .9378  & \textbf{.9484} (27) & .9422  \\
\rowcolor{mygray!70}  &  & & 100 & .9440 & \textbf{.9490} & .9461  \\
&  & & 500 & .9489 & \textbf{.9498} & .9494  \\
 \rowcolor{mygray!70}  &  & & 1000 & .9497 & .9502 & \textbf{.9499}  \\
 &  & & 5000 & .9497 & .9498 & \textbf{.9498}  \\
 \rowcolor{mygray!70}  &  & & 10000 & \textbf{.9503} & .9504 & \textbf{.9503}  \\
  \hline
 & & 0.8 & 50 & .9374  & \textbf{.9577} (4432) & .9402  \\
\rowcolor{mygray!70}  &  & & 100 & .9436 & \textbf{.9514} (20) & .9453  \\
&  & & 500 & .9486 & \textbf{.9498} & .9489  \\
 \rowcolor{mygray!70}  &  & & 1000 & .9496 & \textbf{.9501} & .9498  \\
 &  & & 5000 & \textbf{.9502} & \textbf{.9502} & \textbf{.9502}  \\
 \rowcolor{mygray!70}  &  & & 10000 & \textbf{.9498} & \textbf{.9498} & \textbf{.9498}  \\
\hline
 & 0.8 & 0.8 & 50 & .9322  & .9720 (8901) & \textbf{.9371}  \\
\rowcolor{mygray!70}  &  & & 100 & .9437 & .9569 (45) & \textbf{.9444}  \\
&  & & 500 & .9493 & .9514 & \textbf{.9494}  \\
 \rowcolor{mygray!70}  &  & & 1000 & .9494 & \textbf{.9505} & \textbf{.9495}  \\
 &  & & 5000 & \textbf{.9499} & .9501 & \textbf{.9499}  \\
 \rowcolor{mygray!70}  &  & & 10000 & \textbf{.9501} & \textbf{.9502} & \textbf{.9501}  \\
\bottomrule
\end{tabular}\\
Note: Values closest to 0.95 are displayed in bold.
\label{paired2}
\end{table}

\subsection{Real data analysis}

One significant study that adopts MCC as one of its key evaluation metrics is the research by Le et al. (2021)\cite{le2021transformer}. In their study, a novel biological representation framework aimed at identifying DNA enhancers was proposed. The research utilized an architecture based on Bidirectional Encoder Representations from Transformers (BERT) and 2D Convolutional Neural Networks (CNN).
In the article, they compared the classifier called ``EnhancerPred'' (enhancer predictor) with previous ones, namely EnhancerPred \cite{jia2016enhancerpred}, iEnhancer-2L \cite{liu2016ienhancer}, and iEnhancer-EL \cite{liu2018ienhancer}.
The MCC values for each predictor are summarized in Table \ref{Le}. In the article, they concluded that their predictor demonstrates promising performance compared to the approaches proposed in previous studies.
\begin{table}[hbtp]
\centering
\caption{MCC values for each predictor (Le et al. \cite{le2021transformer}).}
\label{Le}
\begin{tabular}{l|c}
\toprule
Predictor     & MCC   \\ \midrule
EnhancerPred  \ \ \   & 0.480  \\
iEnhancer-2L     & 0.496 \\
iEnhancer-EL      & 0.460  \\
Theirs         & 0.514 \\ \bottomrule
\end{tabular}
\end{table}

By applying the findings from our research, we can calculate the confidence intervals for differences in MCCs. This enables us to determine if there is a statistically significant difference between the MCCs of each classifier, thereby enhancing the credibility of their conclusions.
We encountered challenges in reproducing the results from scratch. Therefore, in this attempt, we aimed to replicate the results by computing $p_{ijk}$ for $i,j,k\in \mathcal{Y}$ (as defined in Eq.~\eqref{a2}) from the numbers of enhancer and non-enhancer samples (200 samples each) along with the sensitivity and specificity of each classifier\footnote{In addition to MCC values, Le et al. \cite{le2021transformer} also presented sensitivity and specificity values. We replicate TP, FP, FN, and TN for each classifier from these values. For more details on these values, see Table \ref{L} in Appendix C or Le et al. \cite{le2021transformer}.}. To uniquely determine the $p_{ijk}$'s, we required two additional independent conditions. Therefore, in this analysis, we examine how the confidence interval of the difference between MCCs changes when varying the values of $p_{001}$ (the probability that the true value is positive, but both classifiers predict negative) and $p_{110}$ (the probability that the true value is negative, but both classifiers predict positive). We utilize the MT method in this context due to its superior performance in scenarios akin to this situation.

\begin{figure}[hbtp]
\begin{center}
\includegraphics[width=150mm]{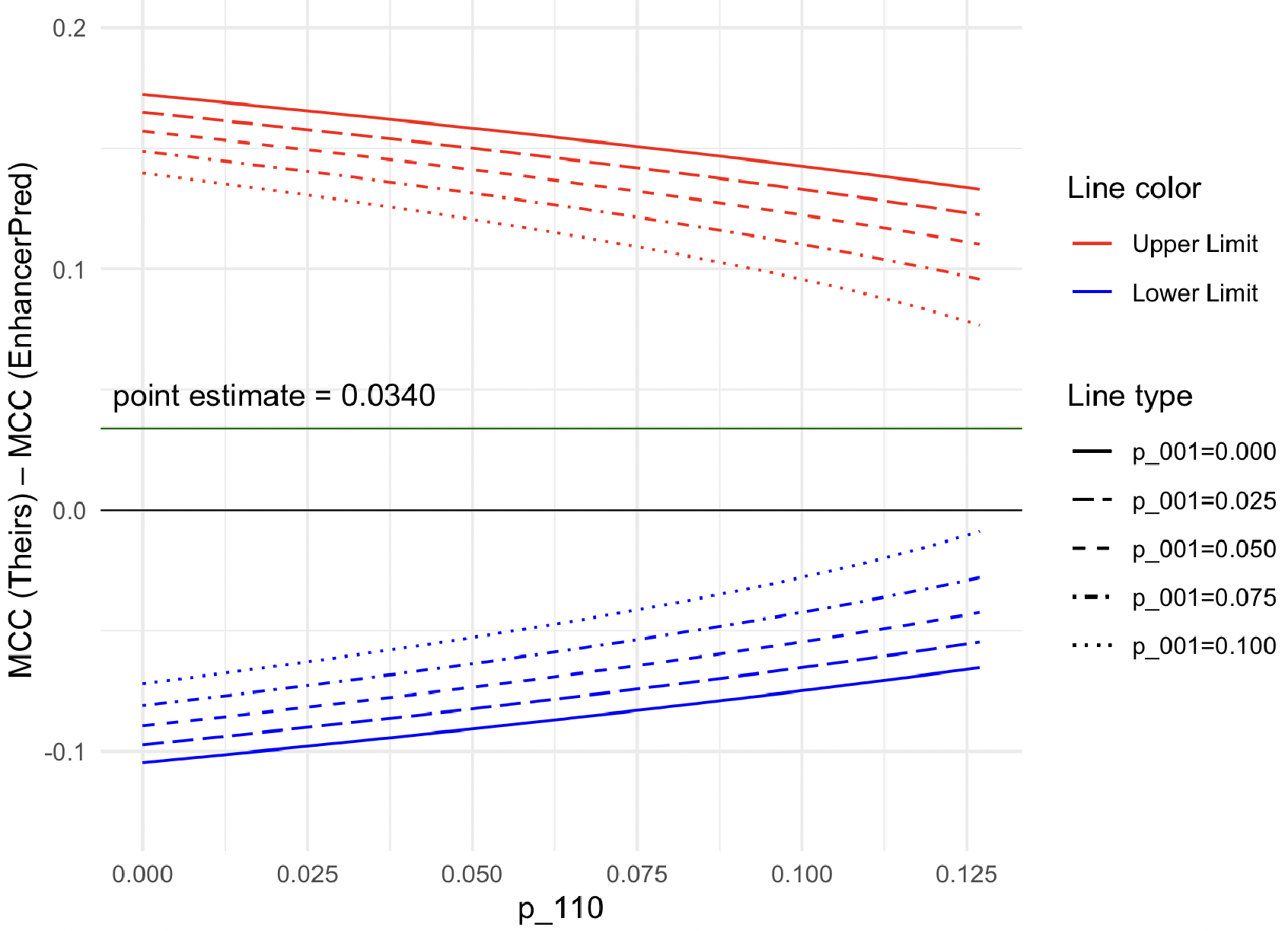}
\caption{The 95\% confidence intervals for difference between MCCs of their predictor and EnhancerPred when varying $p_{001}$ and $p_{110}$.}
\label{pred1}
\end{center}
\end{figure}

First, we compare the performance of their predictor with EnhancerPred. By varying the values of $p_{001}$ and $p_{110}$, we observe the behavior of the confidence intervals. Simple calculations show that the possible values for these parameters in this case are within the following ranges  (refer to Appendix C for details):
\[
0 \leq p_{001} \leq 0.1 \ \ \ \text{and} \ \ \ 0 \leq p_{110} \leq 0.1275.
\]
Figure \ref{pred1} displays the 95\% confidence intervals for the MCC difference between their predictor and EnhancerPred, varying $p_{001}$ in 0.025 increments (indicated by line type) and $p_{110}$ in 0.001 increments ($x$-axis).

Figure \ref{pred1} suggests that, regardless of the values of $p_{001}$ and $p_{110}$, the lower bounds of the confidence intervals for the difference between MCCs are below zero. Thus, at a 5\% significance level and under a two-sided alternative hypothesis, we find no statistically significant difference in performance, as gauged by MCC, between the proposed  predictor and EnhancerPred.

Details of performance comparisons with other existing predictors are provided in Appendix C. Based on these results, for almost all possible combinations of $p_{001}$ and $p_{110}$, we can conclude that there are no statistically significant differences in performance, as gauged by MCC, between their predictor and previous predictors. This conclusion is drawn at a 5\% significance level and under a two-sided alternative hypothesis.
This real data-based analysis demonstrates that utilizing our research findings enables us to extend our conclusions beyond mere comparisons of MCC point estimates.

\section{Conclusion}

Despite the growing need for precise evaluation of classifier performance across various fields, the discussion around the statistical significance of performance metrics, especially of the Matthews Correlation Coefficient (MCC), remains limited. Given the heightened focus on the robustness and reliability of MCC as a crucial performance metric, and acknowledging the growing need for in-depth research on statistical inference of MCC, our study explored and compared various approaches for deriving its asymptotic distribution.

In this paper, we introduced several methods for deriving confidence intervals for both a single MCC and the difference between MCCs in paired designs. Given that MCC represents the Pearson's correlation coefficient between actual and predicted values, we explored several essential ideas from research on the statistical inference of Pearson's correlation coefficient that could be applicable here. For a single MCC, we introduced two methods to construct asymptotic confidence intervals: the first, referred to as the ``Simple method", utilizes the delta method, while the second, referred to as ``Fisher's $z$ method,'' involves applying Fisher's $z$ transformation to MCC to derive the asymptotic confidence interval. To evaluate the difference between MCCs in paired designs, we proposed three approaches: the ``Simple method", which uses the delta method; ``Zou's method", based on the approach proposed by Zou (2007)\cite{zou2007toward} \hspace{-1mm}; and the ``Modified Transformation (MT) method", which applies Fisher's $z$ transformation to the difference between MCCs.

We performed simulations across various scenarios to assess and compare the accuracy of different methods' approximations with finite samples. The results revealed that, in single cases, Fisher's $z$ method outperformed the Simple method. In paired cases, both Zou's method and the MT method were superior to the Simple method. However, decisively concluding which is better between Zou's method and the MT method was not straightforward. Each method displayed unique behavior: the MT method's coverage probability consistently approached the nominal value, while Zou's method tended to produce conservative intervals, particularly with small sample sizes and high MCC values. The simulation results also showed that, in both single and paired situations, balanced scenarios converged more quickly to the nominal coverage probability than imbalanced ones.

Our analysis of real data has illustrated the potential utility of our methods in comparing binary classifiers, highlighting the possible contributions of our research in this field. With a focus on the statistical inference of MCC, this study sought to enhance the accuracy and reliability of classifier evaluations. Our research will serve as a valuable resource in the domain, offering researchers and practitioners effective tools for more accurate and reliable assessments in evaluating and comparing binary classifiers.

Finally, this paper sets the stage for future work to explore additional possibilities, building on the foundation established here. Although our focus has been on binary cases, the evaluation of classifiers in multi-class cases is equally crucial and poses unique challenges. In multi-class contexts, the complexity of MCC increases compared to binary cases. 
As discussed in Takahashi (2022)\cite{takahashi2022confidence} for $\text{F}_1$-scores, both Macro and Micro variants can be considered for MCC. However, a definitive, universally accepted definition of MCC for multi-class settings remains to be established. Our ongoing research is aimed at identifying the most suitable definition of MCC for multi-class situations and providing deeper insights into asymptotic properties of multi-class MCC.

\newpage

\section*{Appendix}
\appendix
\section{Results of differentiation for $\varphi$, $f\circ \varphi$, $\psi$, and $g\circ\psi$  }
This appendix presents the specific results of differentiating the functions discussed in this paper.
For the single cases, we denote $\bm{p}=(p_{11},p_{10},p_{01},p_{00})^{\top}$ and denote the marginal probabilities of the $p_{ij}$'s as
\[
p_{i+} \coloneqq \sum_{j\in \{0,1\}} p_{ij} , \ \ \ p_{+j} \coloneqq \sum_{i\in \{0,1\}} p_{ij}.
\]
Then, the results of the differentiations of the functions $\varphi$ and $f\circ \varphi$ are represented as follows:
\[
\nabla \varphi(\bm{p}) = \left(\frac{\partial \varphi(\bm{p})}{\partial p_{11}},\frac{\partial \varphi(\bm{p})}{\partial p_{10}},\frac{\partial \varphi(\bm{p})}{\partial p_{01}},\frac{\partial \varphi(\bm{p})}{\partial p_{00}}\right)^{\top}
, \ \ \ \ \
\nabla f\circ \varphi(\bm{p}) = f'(\varphi(\bm{p})) \nabla \varphi(\bm{p})= \frac{1}{1-\varphi(\bm{p})^2}\nabla \varphi(\bm{p}),
\]
where
\begin{align*}
\frac{\partial \varphi(\bm{p})}{\partial p_{11}} &= \frac{p_{00}}{\sqrt{p_{1+}p_{+1}p_{+0}p_{0+}}} - \frac{p_{1+}+p_{+1}}{2p_{1+}p_{+1}}\varphi(\bm{p}), \\
\frac{\partial \varphi(\bm{p})}{\partial p_{10}} &= \frac{-p_{01}}{\sqrt{p_{1+}p_{+1}p_{+0}p_{0+}}} - \frac{p_{1+}+p_{+0}}{2p_{1+}p_{+0}}\varphi(\bm{p}), \\
\frac{\partial \varphi(\bm{p})}{\partial p_{01}} &= \frac{-p_{10}}{\sqrt{p_{1+}p_{+1}p_{+0}p_{0+}}} -\frac{p_{+1}+p_{0+}}{2p_{+1}p_{0+}}\varphi(\bm{p}), \\
\frac{\partial \varphi(\bm{p})}{\partial p_{00}}&= \frac{p_{11}}{\sqrt{p_{1+}p_{+1}p_{+0}p_{0+}}} -\frac{p_{0+}+p_{+0}}{2p_{0+}p_{+0}}\varphi(\bm{p}).
\end{align*}

For the paired cases, we denote $\bm{p}=(p_{111},p_{110},p_{101},p_{100},p_{011},p_{010},p_{001},p_{000})^{\top}$ and denote the marginal probabilities of the $p_{ijk}$'s as
\[
p_{i++} \coloneqq \sum_{j,k \in \{0,1\}} p_{ijk}, \ \ \ p_{+j+} \coloneqq \sum_{i,k \in \{0,1\}} p_{ijk}, \ \ \
p_{++k} \coloneqq \sum_{i,j \in \{0,1\}} p_{ijk},
\]
and
\[
p_{ij+} \coloneqq \sum_{k \in \{0,1\}} p_{ijk}, \ \ \ p_{i+k} \coloneqq \sum_{j \in \{0,1\}} p_{ijk}, \ \ \ p_{+jk} \coloneqq \sum_{i \in \{0,1\}} p_{ijk}.
\]
Moreover, we denote the MCCs for each classifier as $\text{MCC}_1$ and $\text{MCC}_2$, respectively:
\begin{align*}
\text{MCC}_1 &\coloneqq \varphi(p_{111}+p_{101},p_{110}+p_{100},p_{011}+p_{001},p_{010}+p_{000}), \\
\text{MCC}_2 &\coloneqq \varphi(p_{111}+p_{011},p_{110}+p_{010},p_{101}+p_{001},p_{100}+p_{000}).
\end{align*}
The denominators for each MCC are denoted as $D_1$ and $D_2$:
\[
D_1 \coloneqq \sqrt{p_{++1}p_{1++}p_{0++}p_{++0}}, \ \ \ \ D_2 \coloneqq \sqrt{p_{++1}p_{+1+}p_{+0+}p_{++0}}.
\]
Then, the results of differentiation of functions $\psi$ and $g\circ \psi$ are represented as follows:
\[
\nabla \psi(\bm{p}) = \left(\frac{\partial \psi(\bm{p})}{\partial p_{111}},\dots,\frac{\partial \psi(\bm{p})}{\partial p_{000}}\right)^{\top},
\ \ \ \ \
\nabla g\circ \psi(\bm{p}) = g'(\psi(\bm{p}))\nabla \psi(\bm{p}) =  \frac{2}{4-\psi(\bm{p})^2}\nabla \psi(\bm{p}),
\]
where $g'$ is the first derivative of $g$ and
\begin{align*}
\frac{\partial \psi(\bm{p})}{\partial p_{111}}  &=  \frac{p_{0+0}}{D_1} - \frac{p_{++1}+p_{1++}}{2p_{++1}p_{1++}} \cdot \text{MCC}_1 -\frac{p_{+00}}{D_2} + \frac{p_{++1}+p_{+1+}}{2p_{++1}p_{+1+}} \cdot \text{MCC}_2 ,\\[5pt]
\frac{\partial \psi(\bm{p})}{\partial p_{110}} &=- \frac{p_{0+1}}{D_1} - \frac{p_{1++}+p_{++0}}{2p_{1++}p_{++0}}\cdot \text{MCC}_1 + \frac{p_{+01}}{D_2} - \frac{p_{+1+}+p_{++0}}{2p_{+1+}p_{++0}} \cdot \text{MCC}_2 ,\\[5pt]
\frac{\partial \psi(\bm{p})}{\partial p_{101}} &=\frac{p_{0+0}}{D_1} - \frac{p_{++1}+p_{1++}}{2p_{++1}p_{1++}} \cdot \text{MCC}_1 + \frac{p_{+10}}{D_2} - \frac{p_{++1}+p_{+0+}}{2p_{++1}p_{+0+}} \cdot \text{MCC}_2 ,\\[5pt]
\frac{\partial \psi(\bm{p})}{\partial p_{100}} &=- \frac{p_{0+1}}{D_1} - \frac{p_{1++}+p_{++0}}{2p_{1++}p_{++0}} \cdot \text{MCC}_1 -\frac{p_{+11}}{D_2} + \frac{p_{+0+}+p_{++0}}{2p_{+0+}p_{++0}} \cdot \text{MCC}_2 ,\\[5pt]
\frac{\partial \psi(\bm{p})}{\partial p_{011}} &=- \frac{p_{1+0}}{D_1} - \frac{p_{++1}+p_{0++}}{2p_{++1}p_{0++}} \cdot \text{MCC}_1 -\frac{p_{+00}}{D_2} + \frac{p_{++1}+p_{+1+}}{2p_{++1}p_{+1+}} \cdot \text{MCC}_2 ,\\[5pt]
\frac{\partial \psi(\bm{p})}{\partial p_{010}} &=  \frac{p_{1+1}}{D_1} - \frac{p_{0++}+p_{++0}}{2p_{0++}p_{++0}} \cdot \text{MCC}_1+ \frac{p_{+01}}{D_2} - \frac{p_{+1+}+p_{++0}}{2p_{+1+}p_{++0}}\cdot \text{MCC}_2 ,\\[5pt]
\frac{\partial \psi(\bm{p})}{\partial p_{001}} &= - \frac{p_{1+0}}{D_1} - \frac{p_{++1}+p_{0++}}{2p_{++1}p_{0++}}\cdot \text{MCC}_1 + \frac{p_{+10}}{D_2} - \frac{p_{++1}+p_{+0+}}{2p_{++1}p_{+0+}}\cdot \text{MCC}_2 , \\[5pt]
\frac{\partial \psi(\bm{p})}{\partial p_{000}} &= \frac{p_{1+1}}{D_1} - \frac{p_{0++}+p_{++0}}{2p_{0++}p_{++0}}\cdot \text{MCC}_1 -\frac{p_{+11}}{D_2} + \frac{p_{+0+}+p_{++0}}{2p_{+0+}p_{++0}}\cdot \text{MCC}_2. \\
\end{align*}

\section{Details in each scenario for simulation study}
In Sections 4.1 and 4.2, simulations were conducted to validate and compare the accuracy of asymptotic approximations of each method in finite samples across a range of scenarios. In this appendix, the specific values of TP, FP, FN, and TN for each scenario in the aforementioned simulation studies are presented. For the single case, the specific values from TP to TN are summarized in Table \ref{scenario1}, while for the paired case, the specific values from $\text{TP}_1$ to $\text{TN}_2$ are compiled in Table \ref{scenario2}.
\begin{table}[htbp]
\renewcommand{\arraystretch}{1.1}
\centering
\caption{True values of TP, FP, FN, and TN in each scenario for single MCC. }
\label{scenario1}
\begin{tabular}{ccll}
\toprule
 $P(Y=1)$ & MCC & &  \\
\midrule
   0.1 & 0.4 & TP=0.0794 & FN=0.0206     \\
  &  & FP=0.1853 & TN=0.7147  \\
  \hline
 \rowcolor{mygray!70}  & 0.6 & TP=0.0890 & FN=0.0110     \\
 \rowcolor{mygray!70}  &  & FP=0.0986 & TN=0.8014  \\
 \hline
 & 0.8 & TP=0.0956 & FN=0.0044     \\
  &  & FP=0.0396 & TN=0.8604  \\
\hline
\rowcolor{mygray!70}  0.5 & 0.4 & TP=0.3500 & FN=0.1500     \\
\rowcolor{mygray!70}  &  & FP=0.1500 & TN=0.3500  \\
\hline
 & 0.6 & TP=0.4000 & FN=0.1000    \\
 &  & FP=0.1000 & TN=0.4000  \\
 \hline
\rowcolor{mygray!70}   & 0.8 & TP=0.4500 & FN=0.0500     \\
\rowcolor{mygray!70}  &  & FP=0.0500 & TN=0.4500 \\
\bottomrule
\end{tabular}
\end{table}

\begin{table}[htbp]
\renewcommand{\arraystretch}{1.1}
\centering
\caption{True values of $\text{TP}_1$ through $\text{TN}_2$ in each scenario in paired design.}
\label{scenario2}
\begin{tabular}{cccllll}
\toprule
$P(Y=1)$ & $\text{MCC}_1$ & $\text{MCC}_2$ &  &  &  &  \\
\midrule
0.1 & 0.4 & 0.4 & $\text{TP}_1=0.0794$  & $\text{FN}_1=0.0206$ & $\text{TP}_2=0.0794$  & $\text{FN}_2=0.0206$  \\
  &  & &  $\text{FP}_1=0.1853$  & $\text{TN}_1=0.7147$ & $\text{FP}_2=0.1853$  & $\text{TN}_2=0.7147$  \\
  \hline
 \rowcolor{mygray!70}  &  & 0.6 & $\text{TP}_1=0.0794$  & $\text{FN}_1=0.0206$ & $\text{TP}_2=0.0890$  & $\text{FN}_2=0.0110$  \\
 \rowcolor{mygray!70} &  & & $\text{FP}_1=0.1853$  & $\text{TN}_1=0.7147$ & $\text{FP}_2=0.0986$  & $\text{TN}_2=0.8014$  \\
\hline
 &  & 0.8 & $\text{TP}_1=0.0794$  & $\text{FN}_1=0.0206$ & $\text{TP}_2=0.0956$  & $\text{FN}_2=0.0044$  \\
  &  & &  $\text{FP}_1=0.1853$  & $\text{TN}_1=0.7147$ & $\text{FP}_2=0.0396$  & $\text{TN}_2=0.8604$  \\
\hline
 \rowcolor{mygray!70}  & 0.6 & 0.6 & $\text{TP}_1=0.0890$  & $\text{FN}_1=0.0110$ & $\text{TP}_2=0.0890$  & $\text{FN}_2=0.0110$  \\
 \rowcolor{mygray!70} &  & & $\text{FP}_1=0.0986$  & $\text{TN}_1=0.8014$ & $\text{FP}_2=0.0986$  & $\text{TN}_2=0.8014$  \\
 \hline
 &  & 0.8 & $\text{TP}_1=0.0890$  & $\text{FN}_1=0.0110$ & $\text{TP}_2=0.0956$  & $\text{FN}_2=0.0044$  \\
  &  & &  $\text{FP}_1=0.0986$  & $\text{TN}_1=0.8014$ & $\text{FP}_2=0.0396$  & $\text{TN}_2=0.8604$  \\
  \hline
 \rowcolor{mygray!70}  & 0.8 & 0.8 & $\text{TP}_1=0.0956$  & $\text{FN}_1=0.0044$ & $\text{TP}_2=0.0956$  & $\text{FN}_2=0.0044$  \\
 \rowcolor{mygray!70} &  & & $\text{FP}_1=0.0396$  & $\text{TN}_1=0.8604$ & $\text{FP}_2=0.0396$  & $\text{TN}_2=0.8604$  \\
\midrule
 0.5 & 0.4 & 0.4 & $\text{TP}_1=0.3500$  & $\text{FN}_1=0.1500$ & $\text{TP}_2=0.3500$  & $\text{FN}_2=0.1500$  \\
  &  & &  $\text{FP}_1=0.1500$  & $\text{TN}_1=0.3500$ & $\text{FP}_2=0.1500$  & $\text{TN}_2=0.3500$  \\
  \hline
 \rowcolor{mygray!70}  &  & 0.6 & $\text{TP}_1=0.3500$  & $\text{FN}_1=0.1500$ & $\text{TP}_2=0.4000$  & $\text{FN}_2=0.1000$  \\
 \rowcolor{mygray!70} &  & & $\text{FP}_1=0.1500$  & $\text{TN}_1=0.3500$ & $\text{FP}_2=0.1000$  & $\text{TN}_2=0.4000$  \\
 \hline
 &  & 0.8 & $\text{TP}_1=0.3500$  & $\text{FN}_1=0.1500$ & $\text{TP}_2=0.4500$  & $\text{FN}_2=0.0500$  \\
  &  & &  $\text{FP}_1=0.1500$  & $\text{TN}_1=0.3500$ & $\text{FP}_2=0.0500$  & $\text{TN}_2=0.4500$  \\
  \hline
   \rowcolor{mygray!70}  & 0.6 & 0.6 & $\text{TP}_1=0.4000$  & $\text{FN}_1=0.1000$ & $\text{TP}_2=0.4000$  & $\text{FN}_2=0.1000$  \\
 \rowcolor{mygray!70} &  & & $\text{FP}_1=0.1000$  & $\text{TN}_1=0.4000$ & $\text{FP}_2=0.1000$  & $\text{TN}_2=0.4000$  \\
  \hline
 &  & 0.8 & $\text{TP}_1=0.4000$  & $\text{FN}_1=0.1000$ & $\text{TP}_2=0.4500$  & $\text{FN}_2=0.0500$  \\
  &  & &  $\text{FP}_1=0.1000$  & $\text{TN}_1=0.4000$ & $\text{FP}_2=0.0500$  & $\text{TN}_2=0.4500$  \\
  \hline
   \rowcolor{mygray!70}  & 0.8 & 0.8 & $\text{TP}_1=0.4500$  & $\text{FN}_1=0.0500$ & $\text{TP}_2=0.4500$  & $\text{FN}_2=0.0500$  \\
   
 \rowcolor{mygray!70} &  & & $\text{FP}_1=0.0500$  & $\text{TN}_1=0.4500$ & $\text{FP}_2=0.0500$  & $\text{TN}_2=0.4500$  \\
\bottomrule
\end{tabular}
\end{table}

\section{Details of real data analysis}
In Section 4.3, we explored the application of our research findings to real-world data, aiming to evaluate their practical applicability and utility.
During this examination, we employed the sensitivity and specificity values for each classifier to reproduce the results reported in Le et al. \cite{le2021transformer}, with these values detailed in Table \ref{L}.
To uniquely determine $p_{ijk}$ for $i,j,k \in \mathcal{Y}$, we need to assign specific values to two of the $p_{ijk}$ variables, taking into account that the sensitivity and specificity for each classifier, along with the sample size and the ratio of positive instances, are provided.
Therefore, we varied $p_{001}$ and $p_{110}$ within their possible ranges and analyzed how the confidence intervals responded to these variations.
Regarding the possible values of $p_{001}$ and $p_{110}$, we can easily verify that
\begin{gather*}
 0 \leq p_{001} \leq P(Y=1) \times(1- \text{max}\{ \text{Sensitivity (Theirs)} ,\ \text{Sensitivity (Previous)} \}), \\
  0 \leq p_{110} \leq P(Y=0) \times (1- \text{max}\{ \text{Specificity (Theirs)} ,\ \text{Specificity (Previous)} \}).
\end{gather*}
That is,
\begin{gather}\label{ine}
\begin{split}
0 \leq p_{001} \leq 0.1, \hspace{2.8cm}\\
  0 \leq p_{110} \leq 0.5\times  (1-\text{Specificity (Previous)}).
  \end{split}
\end{gather}

Figure \ref{pred1} in the main text illustrates the behavior of the confidence intervals for the difference between MCCs of their predictor and EnhancerPred, as $p_{001}$ and $p_{110}$ are varied to satisfy the inequalities in Eq.~\eqref{ine}.
This appendix further extends the comparison to other predictors, as illustrated in Figures 2 and 3.
In almost all combinations of the values for $p_{001}$ and $p_{110}$ in these results, the lower limits of the confidence intervals are below zero. In more detail, Figure 3 shows that the confidence intervals do not contain zero only when $p_{001}$ is close to zero and $p_{110}$ is very close to the upper limit of the possible range. Therefore, it can be concluded that for nearly all combinations of $p_{001}$ and $p_{110}$ values, at a 5\% significance level, no statistical significance can be established between the MCC values of the classifiers.
\vspace{8mm}

\begin{table}[htbp]
\centering
\caption{Comparison of the proposed method and previous methods (Le et al. \cite{le2021transformer}).}
\label{L}
\begin{tabular}{l|cccc}
\toprule
Predictor     & Sensitivity (\%) & Specificity (\%) & Accuracy (\%) & MCC   \\ \midrule
EnhancerPred  & 73.5             & 74.5             & 74            & 0.48  \\
iEnhancer-2L  & 71               & 78.5             & 74.75         & 0.496 \\
iEnhancer-EL  & 71               & 75               & 73            & 0.46  \\
Theirs          & 80               & 71.2             & 75.6          & 0.514 \\ \bottomrule
\end{tabular}
\end{table}

\begin{figure}[htbp]
\begin{center}
\includegraphics[width=140mm]{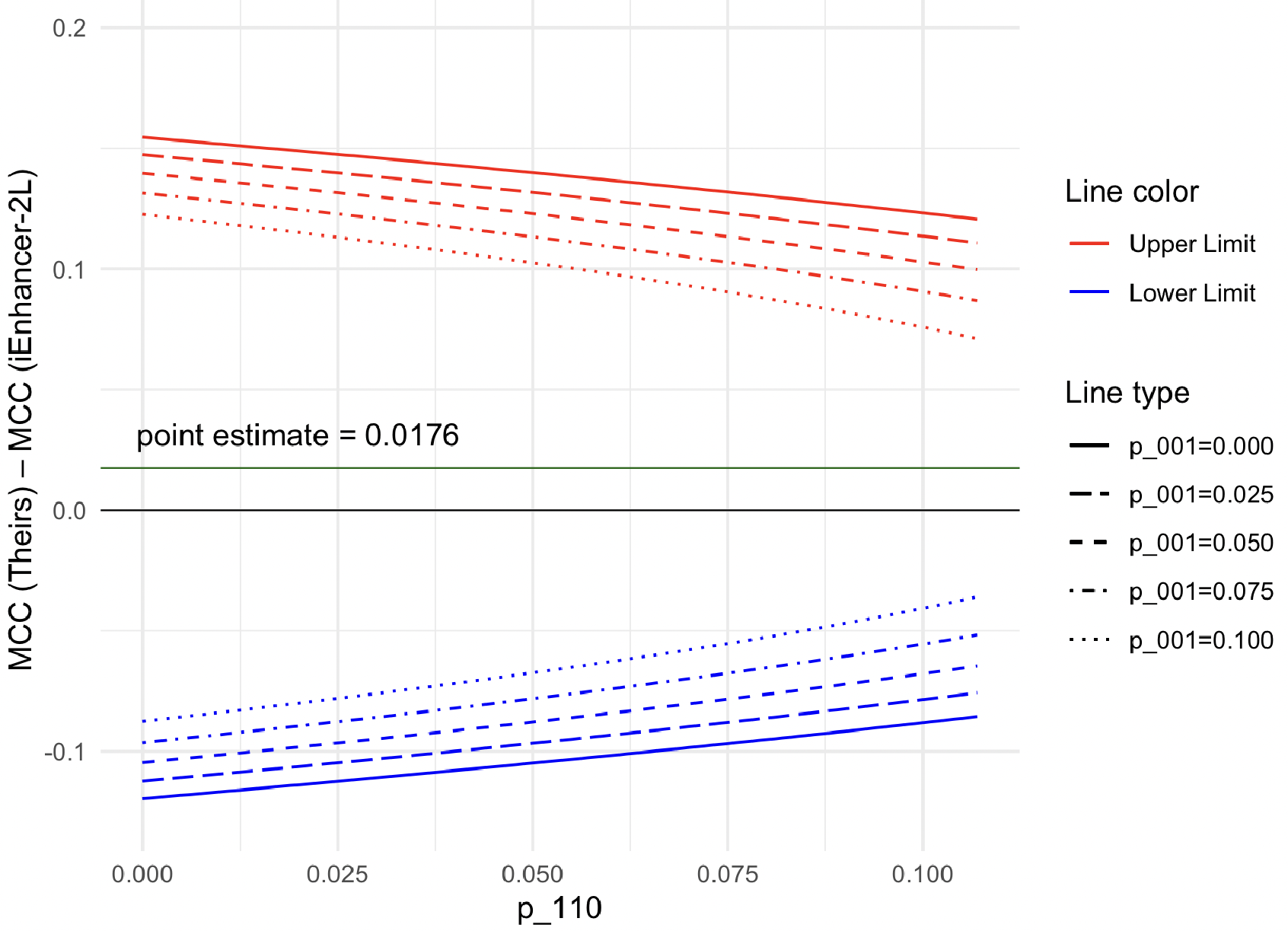}
\caption{The 95\% confidence interval for difference between MCCs of their predictor and iEnhancer-2L when varying $p_{001}$ and $p_{110}$.}
\end{center}
\end{figure}

\begin{figure}[htbp]
\begin{center}
\includegraphics[width=140mm]{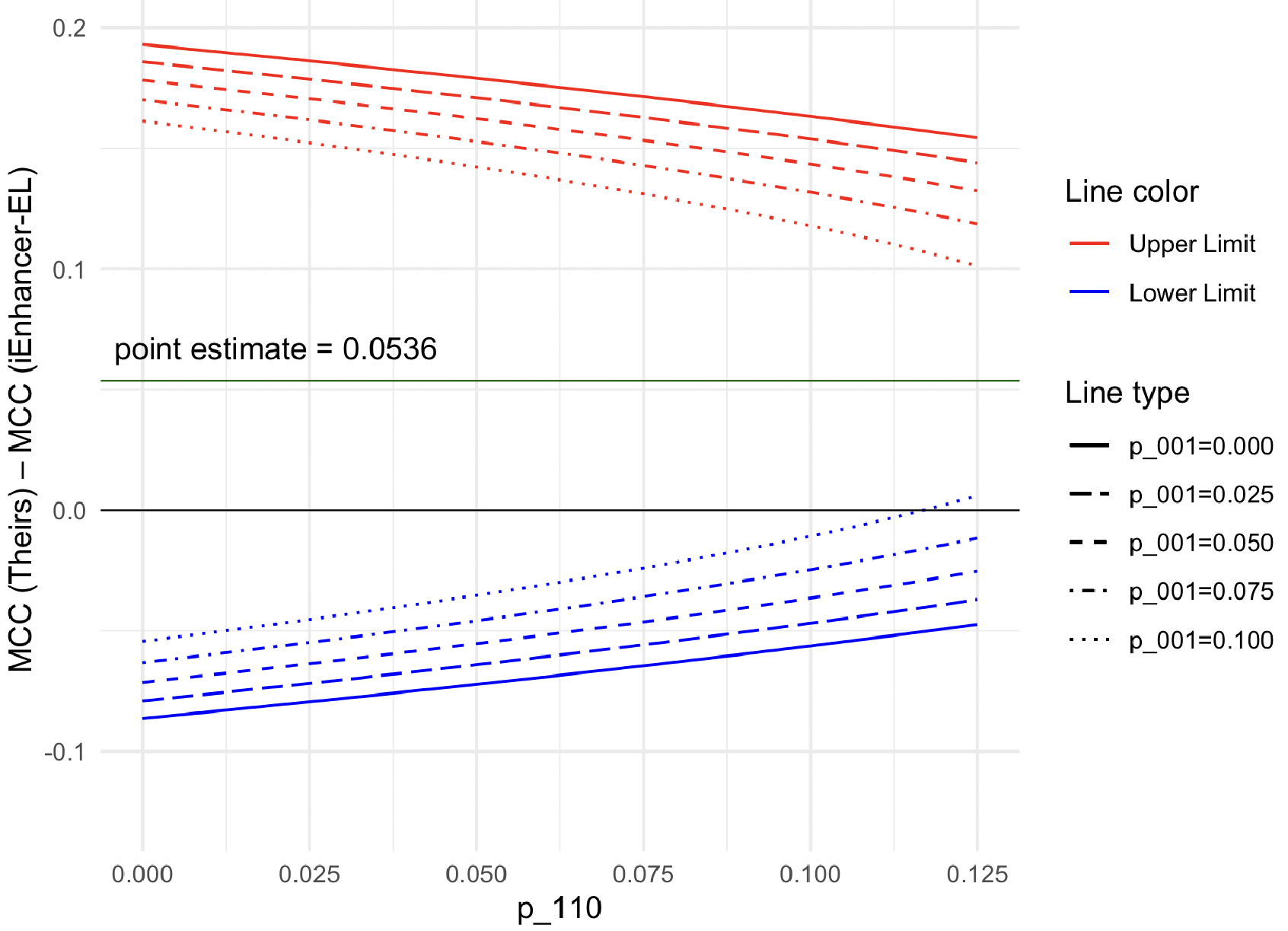}
\caption{The 95\% confidence interval for difference between MCCs of their predictor and iEnhancer-EL when varying $p_{001}$ and $p_{110}$.}
\end{center}
\end{figure}

\end{document}